\documentclass[%
 aip,
 amsmath,amssymb,
 reprint,%
]{revtex4-1}

\usepackage{graphicx}
\usepackage{dcolumn}
\usepackage{bm}
\usepackage{adjustbox}
\usepackage[utf8]{inputenc}
\usepackage[T1]{fontenc}
\usepackage{mathptmx}
\usepackage{xcolor}
\usepackage{hyperref}
\hypersetup{
	colorlinks   = true,
	citecolor    = blue,
	linkcolor = blue,
    urlcolor = blue
}

\begin{document}

\preprint{AIP/123-QED}

\title{The intrinsic ferromagnetism of two-dimensional (2D) MnO$_2$ revisited: A many-body Quantum Monte Carlo and DFT+U study}

\author{Daniel Wines}
\affiliation{%
Department of Physics, University of Maryland Baltimore County, Baltimore MD 21250
}%

\author{Kayahan Saritas}%

\affiliation{ 
Department of Applied Physics, Yale University, New Haven CT 06520
}%

\author{Can Ataca}
 \email{ataca@umbc.edu}
\affiliation{%
Department of Physics, University of Maryland Baltimore County, Baltimore MD 21250
}%

\date{\today}

\begin{abstract}

Monolayer MnO$_2$ is one of the few predicted two-dimensional (2D) ferromagnets that has been experimentally synthesized and is commercially available. The Mermin-Wagner theorem states that magnetic order in a 2D material cannot persist unless magnetic anisotropy (MA) is present and perpendicular to the plane, which permits a finite critical temperature. Previous computational studies have predicted the magnetic ordering and Curie temperature of 2D MnO$_2$ with DFT+U (Density Funtional Theory + Hubbard U correction), with the results having a strong dependence on the Hubbard U parameter. Diffusion Monte Carlo (DMC) is a correlated electronic structure method that has had demonstrated success for the electronic and magnetic properties of a variety of 2D and bulk systems since it has a weaker dependence on the starting Hubbard parameter and density functional. In this study, we used DMC and DFT+U to calculate the magnetic properties of monolayer MnO$_2$. We found that the ferromagnetic ordering is more favorable than antiferromagnetic and determined a statistical bound on the magnetic exchange parameter ($J$). In addition, we performed spin-orbit MA energy calculations using DFT+U and using our DMC and DFT+U parameters along with the analytical model of Torelli and Olsen, we estimated an upper bound of 28.8 K for the critical temperature of MnO$_2$. These QMC results intend to serve as an accurate theoretical benchmark, necessary for the realization and development of future 2D magnetic devices. These results also demonstrate the need for accurate methodologies to predict magnetic properties of correlated 2D materials.

\end{abstract}

\maketitle


\section{\label{sec:intro}Introduction}

In the past decade, the search for two-dimensional (2D) magnets, especially ferromagnetic materials, has come to the forefront of the materials science community. With the experimental realization of CrI$_3$, which has a measured Curie temperature of 45 K \cite{cri3}, interest in identifying similar ferromagnetic materials has increased. In addition to 2D CrI$_3$, it has been demonstrated that ferromagnetic order persists down to the bilayer limit in Cr$_2$Ge$_2$Te$_6$ \cite{crgete} and room temperature magnetic order has been observed for monolayer VSe$_2$ on a van der Waals substrate \cite{vse2}. In addition, computational studies have predicted magnetic ordering in a variety of 2D materials such as K$_2$CuF$_4$ \cite{PhysRevB.88.201402}, the family of MPX$_3$ (where M is 3d transition metal atom, X is group VI atom) \cite{PhysRevB.94.184428}, $\alpha$-RuCl$_3$ \cite{C7CP07953B}, RuBr$_3$ and RuI$_3$ \cite{ERSAN2019111}, and several others. 

Monolayer MnO$_2$ is a 2D layered semiconducting transition metal oxide material that has been reliably experimentally synthesized and studied extensively with computational methods \cite{mno2-exp,PhysRevB.93.045132,Rong_2019,irradiation,C1CP20634F,https://doi.org/10.1002/adma.201602281,ataca-mx2,fm-mno2}. In a previous work by Kan et al. \cite{fm-mno2}, the ferromagnetic (FM) ordering of 2D MnO$_2$ was predicted to be more favorable than the antiferromagnetic (AFM) ordering with DFT+U, using a Hubbard U correction from previous literature (U = 3.9 eV \cite{PhysRevB.73.195107}). The magnetic exchange parameters ($J$) were extracted from these DFT+U calculations and then a magnetic coupling Hamiltonian based on the Ising model was constructed using $J$. This Hamiltonian was then used for classical Monte Carlo simulations to obtain the Curie temperature, which was calculated to be 140 K \cite{fm-mno2}. Although these results are promising, there are certain aspects of the calculations that can be revisited with more sophisticated techniques. 

The Mermin-Wagner theorem \cite{PhysRevLett.17.1133} states that magnetic order in a 2D material cannot persist unless magnetic anisotropy (MA) is present and perpendicular to the plane, which permits a finite critical temperature. Therefore to obtain an accurate value for the critical temperature, the magnetic anisotropy energies (MAE) should be determined by performing spin-orbit calculations. In addition, previous results (and benchmarking results presented in this work) are heavily influenced by the choice of Hubbard parameter (U) \cite{fm-mno2}. Due to this, a method that has less of a dependence on the U parameter that can explicitly capture the electron correlation effects that drive magnetic ordering is desirable. By determining the MAE and $J$ parameters of a 2D system with improved accuracy, analytical models such as the one derived by Torelli and Olsen \cite{Torelli_2018} can be used in conjunction with these ab-initio parameters to estimate the critical temperature. In addition, the realization of 2D magnetic device fabrication can be expedited by using more accurate many-body methods. 

Diffusion Monte Carlo (DMC) \cite{RevModPhys.73.33} is a correlated electronic structure method that has had demonstrated success for the electronic and magnetic properties of a variety of 2D and bulk systems \cite{ataca_qmc,PhysRevB.95.081301,PhysRevB.96.075431,PhysRevMaterials.3.124414,Luo_2016,C6CP02067D,doi:10.1063/1.4919242,PhysRevB.98.155130,doi:10.1063/5.0022814,bennett2021origin,PhysRevB.103.205206,PhysRevX.4.031003,PhysRevB.94.035108,PhysRevMaterials.2.085801,PhysRevMaterials.5.024002,doi:10.1063/5.0023223,wines2021pathway,bilayer-phos,PhysRevX.9.011018,PhysRevLett.115.115501,doi:10.1063/1.5026120,phosphors,PhysRevMaterials.1.065408,staros}.  This method has a weaker dependence on the starting Hubbard parameter and density functional and can successfully achieve results with an accuracy beyond the mean field approximation \cite{RevModPhys.73.33}. 
For example, DMC has succesfully been used to calculate the spin superexchange in the correlated cuprate Ca$_2$CuO$_3$ \cite{PhysRevX.4.031003}, has been used to successfully predict the magnetic structure in FeSe when DFT methods disagreed \cite{PhysRevB.94.035108}, and has been applied to bulk polymorphs of MnO$_2$ to achieve band gap and lattice constant values in excellent agreement with experiment \cite{PhysRevMaterials.2.085801} and has been applied to study the excitation energies of Mn$^{4+}$ doped phosphors \cite{phosphors} (both Mn-based studies used the same RRKJ pseudopotentials we used in our work \cite{PhysRevMaterials.2.085801,phosphors}). 

In this work, we used DMC in conjunction with DFT+U to determine the the magnetic properties of 2D MnO$_2$. We found that the FM ordering is more favorable than AFM and calculated a statistical bound on $J$. We also performed spin-orbit MAE calculations using DFT+U and used these parameters in conjunction with an analytical model to estimate the critical temperature of MnO$_2$, taking into account magnetic exchange and magnetic anisotropy. In Section \ref{sec:methods} we outline the DFT, DFT+U and DMC calculation details, along with additional details of our critical temperature estimation. In Section \ref{sec:results} we present in detail our DMC calculated magnetic exchange energies and magnetic exchange parameters and benchmark with various DFT functionals and different Hubbard parameters, our calculated MAE with DFT+U (spin-orbit calculations) and our estimated critical temperatures with respect to U. Finally we provide concluding remarks and future perspectives in Section \ref{sec:conclusion}.

\section{\label{sec:methods}Computational Methods}

Reference DFT calculations were performed using the VASP code with projector augmented wave (PAW) pseudopotentials \cite{PhysRevB.54.11169,PhysRevB.59.1758}. For these VASP benchmarking calculations, the Perdew-Burke-Ernzerhof (PBE)\cite{PhysRevLett.77.3865}, local density approximation (LDA)\cite{PhysRev.136.B864} and strongly constrained and appropriately normed (SCAN)\cite{PhysRevLett.115.036402} meta-GGA functionals were used. For benchmarking purposes, these DFT calculations (PBE, LDA, SCAN) were additionally performed with the added Hubbard correction (U) \cite{PhysRevB.57.1505} to treat the on-site Coulomb interaction of the $3d$ orbitals of the Mn atoms, where various U values were tested. In order to make a more systematic comparison to DFT+U, we also performed calculations with the screened hybrid HSE06 functional, which is formed by mixing 75$\%$ of the PBE exchange with 25$\%$ of the Fock
exchange and 100$\%$ of the correlation energy from PBE \cite{doi:10.1063/1.1564060}. At least 20 \AA\space of vacuum was given between periodic layers of MnO$_2$ in the $c$-direction. We used a kinetic energy cutoff of 600 eV and a 12x12x1 reciprocal grid for the FM/AFM supercell (12 atoms, 2x2x1 of the primitive cell). The number of k-points were appropriately scaled with the supercell size. In order to determine the magnetic anisotropy energies (MAE), spin-orbit DFT+U calculations were carried out using PBE and PAW potentials (VASP code) for the FM and AFM states of MnO$_2$. The MAE is determined by performing two spin-orbit calculations total energy calculations, one calculation where the spins are oriented in the off-plane direction ($z$ in our case) and one calculation where the easy axis is rotated 90$^{\circ}$ ($x$ in our case). 

For our QMC simulations, we used DFT-PBE to generate the trial wavefunction for subsequent fixed-node DMC calculations. For our DFT calculations within the QMC workflow, the Quantum Espresso (QE) \cite{Giannozzi_2009} code was used. In addition, the trial wavefunction was created for the FM and AFM configurations of 2D MnO$_2$ using various U values. This was done in order to variationally determine the optimal nodal surface (U value that yields the lowest total energy). For Mn and O atoms, we used hard norm-conserving RRKJ (OPT) pseudopotentials \cite{PhysRevB.93.075143}. These potentials have been thoroughly tested in previous DMC works for Mn and O-based materials \cite{PhysRevB.93.075143,PhysRevMaterials.2.085801}. For these pseudopotentials, we used a kinetic energy cutoff of 300 Ry. We tested the reciprocal grid size at the DFT level and determined that for a 12 atom supercell (2x2x1 of the primitive cell), a k-grid of 3x3x1 was sufficient (see Supporting Information, Fig. S1).

Variational Monte Carlo (VMC) and DMC \cite{RevModPhys.73.33,Needs_2009} calculations were carried out using the QMCPACK \cite{Kim_2018,doi:10.1063/5.0004860} code after the trial wavefunction was generated using DFT. VMC calculations serve as the intermediate steps between the DMC and DFT calculations, where the single determinant DFT wavefunction is converted into a many-body wavefunction, by use of the Jastrow parameters \cite{PhysRev.34.1293,PhysRev.98.1479}. Jastrow parameters assist in modeling the electron correlation and subsequently reduce the uncertainty in the DMC calculations \cite{PhysRevLett.94.150201,doi:10.1063/1.460849}. Up to three-body Jastrow \cite{PhysRevB.70.235119} correlation functions were included. The linear method \cite{PhysRevLett.98.110201} was used to minimize the variance and energy respectively of the VMC energies. The cost function of the variance optimization is 100$\%$ variance minimization and the cost function of the energy optimization is split as 95$\%$ energy minimization and 5$\%$ variance minimization, which has been demonstrated to reduce the uncertainty for DMC results \cite{PhysRevLett.94.150201}. The automated DFT-VMC-DMC workflows were generated using the Nexus \cite{nexus} software suite. DMC calculations were performed at supercells sizes of 36 atoms for the FM and AFM configurations of MnO$_2$ and results for magnetic exchange energy were compared to smaller supercell sizes to demonstrate the convergence of supercell size. This comparison is illustrated in Fig. S2, where we observe the FM and AFM energy difference of the 18, 24 and 36 atom cells to be statistically identical to each other and statistically identical to the infinite size extrapolated value, which demonstrates that the finite size convergence for the quantities of interest in this study can be obtained using a 36 atom cell. The locality approximation \cite{doi:10.1063/1.460849} was used to evaluate the nonlocal part of the pseudopotentials in DMC and a timestep of 0.01 Ha$^{-1}$ was used for all DMC simulations (as tested in our previous work for bulk MnO$_2$ polymorphs \cite{PhysRevMaterials.2.085801}).

We extracted the total charge density and spin density from our DMC calculations. The extracted spin density ($\rho_s$) is the difference between the spin-up contribution to the total charge density and the spin-down contribution to the total charge density ($\rho_s = \rho_{up}-\rho_{down}$). An extrapolation scheme was used on the DMC charge densities to eliminate the bias that occurs from using a mixed estimator. Due to the fact that the charge density estimator does not commute with the fixed-node Hamiltonian, the DMC charge density we obtained is a mixed estimator between the pure fixed-node DMC and VMC densities. The extrapolation formula takes the following form\cite{RevModPhys.73.33}:

\begin{equation} \label{rho1}
\rho_1 =2\rho_{\textrm{DMC}}-\rho_{\textrm{VMC}}+\mathcal{O}[(\Phi-\Psi_{\textrm{T}})^2]
\end{equation}
where $\rho_{\textrm{DMC}}$ and $\rho_{\textrm{VMC}}$ are the DMC and VMC charge densities respectively. $\Phi$ is the trial wavefunction from the DMC Hamiltonian and $\Psi_{\textrm{T}}$ is the trial wavefunction from VMC. 

We went one step further and integrated the DFT and DMC spin densities up to a cutoff radius $r_{cut}$ (which we define as 0.9 \AA\space, due to the fact that it is half of the Mn-O bond distance in 2D MnO$_2$) to estimate the site-averaged atomic magnetic moment per Mn and O. In order to obtain these magnetic moments per atom ($M_A$), we sum over the spherically interpolated spin densities: 

\begin{equation} \label{M_a}
M_A = 4\pi \int_0^{r_{cut}}r^2 \rho_s(r)dr \approx 4\pi \sum_{i=0}^{r_{cut}/\Delta r}r_i^2 \rho_s(r_i)\Delta r
\end{equation}
where $r_i$ is the distance from the center of the atom to a given point on the grid and $\Delta r$ is the radial grid size.

In order to calculate the critical temperature of 2D MnO$_2$, we used the method outlined in Torelli and Olsen \cite{Torelli_2018}, which derived a simple expression for the critical temperature of 2D magnetic materials by fitting classical Monte Carlo results for different lattice structures. This expression is a function of the ab-initio calculated MAE and magnetic exchange constants. We decided to use the Torelli and Olsen method (based on classical Monte Carlo fitting) instead of the method of Lado and Fernandez-Rossier (based on spin wave theory) \cite{Lado_2017} due to the fact that the Torelli and Olsen method has been shown to estimate a critical temperature with experimental accuracy for CrI$_3$\cite{Torelli_2018} where the Lado and Fernandez-Rossier method has been shown to underestimate the critical temperature of CrI$_3$ by 20$\%$ \cite{Lado_2017}.  From our DMC calculated magnetic exchange constants and DFT+U obtained MAE, we were able to obtain an upper and lower bound on the critical temperature.

\section{\label{sec:results}Results and Discussion}

\begin{figure*}
\begin{center}
\includegraphics[width=14.5cm]{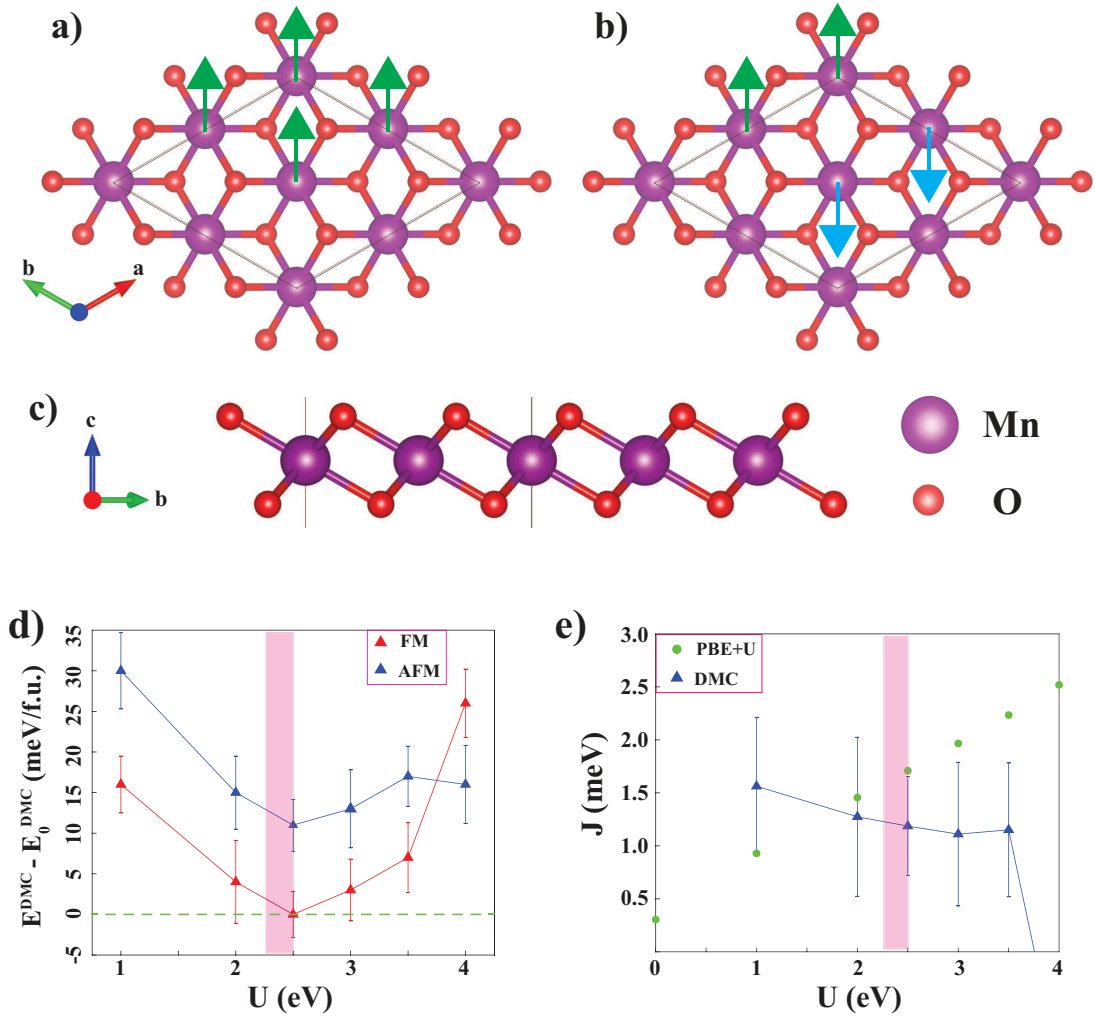}
\caption{The top (a-b) and side (c) view of the atomic structure of monolayer MnO$_2$. The ferromagnetic (FM) ordering is depicted in a) while the antiferromagnetic (AFM)  ordering is depicted in b) where green arrows represent spin-up and blue arrows represent spin-down. d) The DMC calculated total energies of a 36 atom supercell (normalized per formula unit (f.u.), 3 atoms) of the ferromagnetic (red) and antiferromagnetic (blue) states of 2D MnO$_2$ calculated as a function of the U parameter used to variationally determine the optimal trial wave function. For convenience of presentation, the DMC energies are shifted by E$_0^{\textrm{DMC}}$ (the lowest DMC energy obtained for the FM ordering at U = 2.5 eV) and the green dashed line is drawn at E$^{\textrm{DMC}}_0$. e) The nearest neighbor magnetic exchange parameter $J$ obtained from DMC (blue triangle) and PBE+U (green circle) calculations as a function of U value, using RRKJ pseudopotentials. Due to the fact that $J$ is negative (-1.1(7) meV) for U = 4 eV, the data point is out of the range of e). The magenta rectangle in d) - e) represents the optimal fitted U value of 2.4(1) eV.   }
\label{geofig}
\end{center}
\end{figure*}

Two dynamically stable phases, H (the trigonal prismatic phase (2H)-hexagonal honeycombs) and T (octahedral phase (1T)-centered honeycombs) of monolayer MnO$_2$ have been predicted stable from phonon and ab-initio molecular dynamics simulations \cite{ataca-mx2}. In this work, we will focus on the T-phase of 2D MnO$_2$, which has been predicted to be semiconducting and magnetic, from previous studies and our own work \cite{ataca-mx2}. Although the unit cell of MnO$_2$ can be constructed from 3 atoms (one Mn and two O), in order to study the FM and AFM states separately, larger supercells must be constructed to avoid periodic interactions of the Mn spins of the AFM state. We constructed 2x2x1 (12 atoms) and 2x1x1 (6 atoms) supercells to combat this for various DFT+U calculations. Unless otherwise noted, we normalized all of our calculated quantities to the 2x2x1 (12 atom) supercell.  

The bulk counterpart of monolayer MnO$_2$ (in this work monolayer MnO$_2$ will solely refer to T-phase) is layered $\delta$-MnO$_2$, which is among several polymorphs of bulk MnO$_2$ ($\beta$, R, $\alpha$, $\gamma$, $\lambda$) \cite{PhysRevB.93.045132}. The top and side view of the atomic structure of the 2x2x1 supercell of 2D MnO$_2$ is given in Fig. \ref{geofig} a - c), for the FM and AFM ordering. In monolayer MnO$_2$, each Mn atom is bonded to six O atoms in an octahedral configuration. As reported in previous literature\cite{fm-mno2} and from our own spin polarized DFT calculations, the FM state has a net magnetic moment of 3 $\mu_B$ per unit cell arising from Mn atoms. This can be explained by the octahedral coordination, where the $d$ orbitals are split into t$_{2\textrm{g}}$ (d$_{xy}$, d$_{yz}$, d$_{xz}$) and e$_\textrm{g}^{*}$ (d$_{z^2}$, d$_{x^2-y^2}$) orbitals, where t$_{2\textrm{g}}$ orbitals have lower energy. Since each octahedrally coordinated Mn atom has a configuration of 3d$^3$, and the three t$_{2\textrm{g}}$ orbitals are singly occupied with parallel spins (Hund's rule), the unit cell has a resulting magnetic moment of 3 $\mu_B$.

\begin{table}[]
\caption{\label{tab:vasp1}
Benchmarking data obtained with various functionals (PBE, SCAN, LDA) and different U values using the VASP code and PAW pseudopotentials for monolayer MnO$_2$. For each functional/U value, the geometry of 2D MnO$_2$ was fully relaxed separately for the FM and AFM states (12 atom supercell, 2x2x1 of the primitive cell). The first column depicts the functionals and U value used, the second column depicts the energy differences between the optimized FM state geometry and the optimized AFM geometry, the third column depicts the optimal lattice constant of the FM state and the fourth column depicts the optimal lattice constant of the AFM state.  }
\begin{tabular}{lrrr}
\hline
\hline
Functional & \multicolumn{1}{r}{E$_{\textrm{FM}}$-E$_{\textrm{AFM}}$} & \multicolumn{1}{r}{FM} & \multicolumn{1}{r}{AFM} \\
& (meV) & a=b (\AA) & a=b (\AA) \\
\hline
PBE, U=0        & -8.2                        & 5.772                 & 5.749                  \\
PBE, U=2      & -36.6                          & 5.806                 & 5.789                  \\
PBE, U=3.5    & -55.0                        & 5.838                 & 5.817                  \\
SCAN, U=0       & -3.1                           & 5.696                 & 5.685                  \\
SCAN, U=2     & -16.0                           & 5.723                 & 5.712                  \\
SCAN, U=3.5   & -28.7                          & 5.754                 & 5.735                  \\
LDA, U=0        & 18.0                            & 5.636                 & 5.615                  \\
LDA,U=2      & -25.4                          & 5.664                  & 5.650                  \\
LDA, U=3.5    & -44.1                          & 5.685                  & 5.671                 \\
\hline
\hline
\end{tabular}
\end{table}

Prior to any DMC calculations, we benchmarked the optimal geometry and energy of the FM and AFM states of monolayer MnO$_2$ using the VASP code (PAW potentials) with various density functionals and U corrections. The goal of these calculations was to determine whether the preferred magnetic ordering (if ground state was FM or AFM) has a strong dependence on the density functional, geometry or U correction. In Table \ref{tab:vasp1}, we report the energy difference between the relaxed FM and AFM configurations, and the optimal lattice constant for the FM and AFM configurations for PBE, SCAN and LDA using U values of 0, 2 and 3.5 eV (for the 2x2x1, 12 atom supercell). We chose these values of U to obtain a robust sampling of results close to (and surrounding) results previously reported in literature \cite{fm-mno2,PhysRevB.73.195107}. The energy difference between the FM and AFM state is an important quantity because it can give insight to which magnetic state is more favorable in nature. In addition, the magnetic exchange parameters are directly calculated from the E$_{\textrm{FM}}$-E$_{\textrm{AFM}}$ energy. 

In Table \ref{tab:vasp1}, we observe that the E$_{\textrm{FM}}$-E$_{\textrm{AFM}}$ energy is negative (FM favorable) for PBE and SCAN (U = 0, 2, 3.5 eV), LDA (U = 2.5, 3 eV), and positive (AFM favorable) for LDA (U = 0 eV). As the U value is increased, the E$_{\textrm{FM}}$-E$_{\textrm{AFM}}$ energy becomes more negative for each functional. In addition to the strong functional and U dependence on the E$_{\textrm{FM}}$-E$_{\textrm{AFM}}$ energy, there is also strong dependence on the optimal lattice constant, as seen in the last two columns of Table \ref{tab:vasp1}. With comparison to the experimental value of 5.69(1) \AA\space \cite{PhysRevB.93.045132,mno2-geo1,mno2-2,https://doi.org/10.1002/adma.200306592} for the bulk/thin film layered $\delta$-MnO$_2$, the relaxed lattice constant of the FM ordered monolayer structure calculated with SCAN (U = 0 eV) is the closest to the experimental value. It is also noted that lattice constants of AFM ordering are slightly smaller than FM ordering for every functional and U values. As expected, PBE tends to overestimate the lattice constant, LDA tends to underestimate, and SCAN (meta-GGA) can achieve highly accurate lattice constants, as demonstrated in previous works \cite{doi:10.1063/5.0023223,buda}. For this reason, we used the fixed lattice constant and atomic positions obtained with SCAN (for FM ordering) for all subsequent DMC and DFT+U spin-orbit calculations (discussed later in this section). 

\begin{table}[]
\caption{\label{tab:vasp2}
Benchmarking data obtained with various functionals (PBE, SCAN, LDA) using the VASP code and PAW pseudopotentials for monolayer MnO$_2$. To investigate the geometry dependence of the FM/AFM energies we took the relaxed geometries obtained with each functional and calculated the energy with a different functional (for the 12 atom supercell, 2x2x1 of the primitive cell). The first column displays which geometry was used, the second column depicts the functional used to calculate the energies, the third column depicts the energy differences between the FM state and the AFM state. }
\begin{tabular}{llr}
\hline
\hline
Geometry & Functional & \multicolumn{1}{l}{E$_{\textrm{FM}}$-E$_{\textrm{AFM}}$ } \\
& & (meV) \\
\hline
PBE      & PBE        & -8.2                        \\
SCAN     & PBE        & -4.2                         \\
LDA      & PBE        & -13.0                        \\
\hline
PBE      & SCAN       & 3.1                         \\
SCAN     & SCAN       & -3.1                           \\
LDA      & SCAN       & -10.7   \\
\hline
PBE      & LDA        & 13.9                           \\
SCAN     & LDA        & 11.2                          \\
LDA      & LDA        & 18.0                            \\

\hline
\hline
\end{tabular}
\end{table}

To investigate further whether the E$_{\textrm{FM}}$-E$_{\textrm{AFM}}$ energy is functional dependent or geometry dependent, we decided to fix the FM geometry obtained with one functional, and calculate the E$_{\textrm{FM}}$-E$_{\textrm{AFM}}$ energy with another functional. These results are tabulated in Table \ref{tab:vasp2}. Here we see that regardless of the functional used to optimize the geometry, the energies for each of the three geometries are within a few meV of each other when calculated with PBE, SCAN and LDA respectively. Most evidently, LDA is predicting the AFM state to be more favorable regardless of which geometry is used. Based on the benchmarking DFT (PAW) data presented in Table \ref{tab:vasp1} and Table \ref{tab:vasp2}, we observe that the E$_{\textrm{FM}}$-E$_{\textrm{AFM}}$ energy is most dependent on the functional and U parameter used and not the optimal geometry. Due to the weak dependence on geometry, we decided to use DMC to calculate the energy of the FM and AFM state of 2D MnO$_2$ separately, since DMC has a weak dependence on the starting wavefunction, density functional and Hubbard parameter.

\begin{figure}
\begin{center}
\includegraphics[width=8cm]{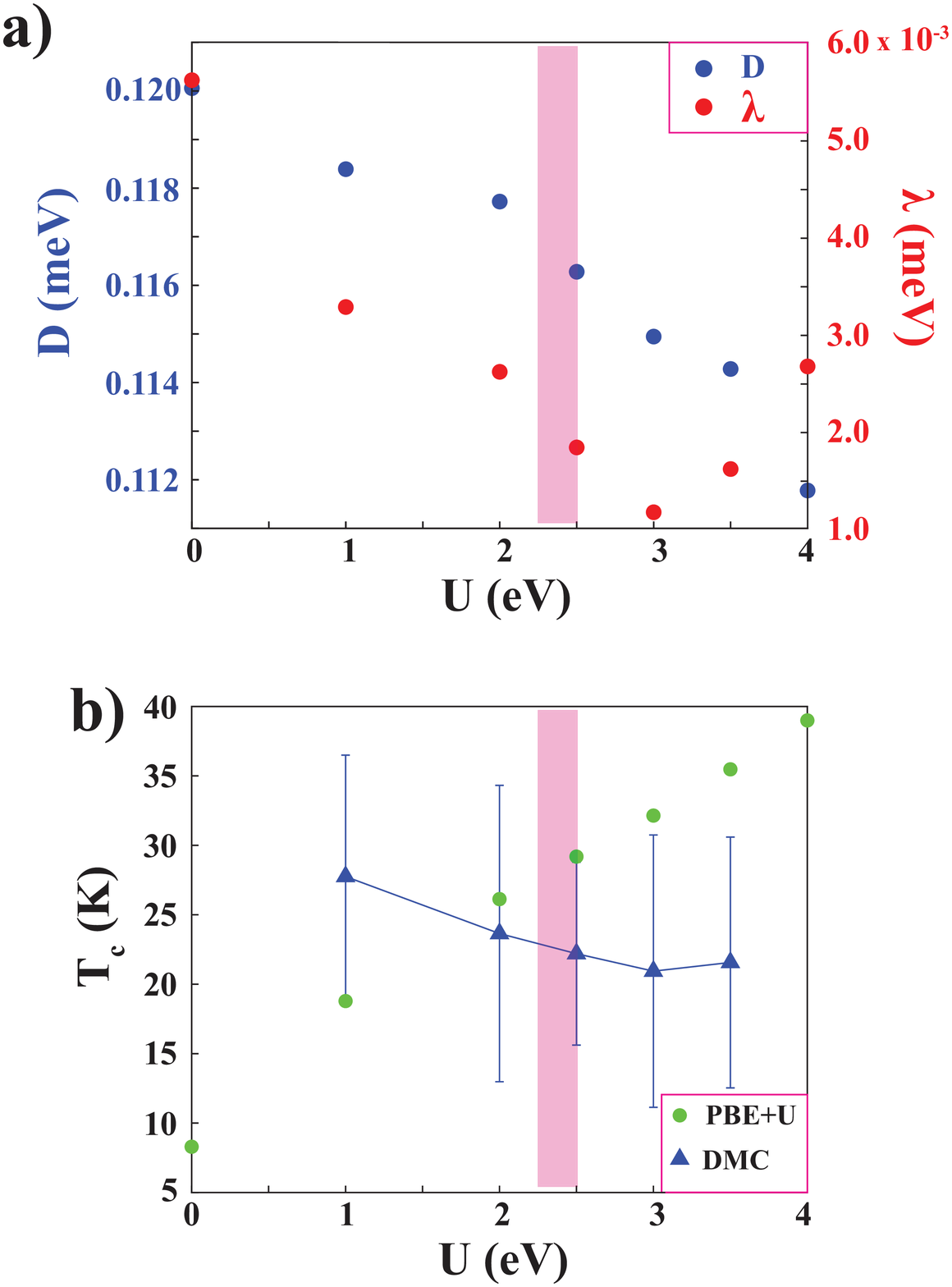}
\caption{a) The single ion anisotropy D (blue circle) and the anisotropic exchange $\lambda$ (red circle) as a function of U, obtained from spin-orbit PBE+U calculations using PAW pseudopotentials. b) The estimated critical temperature obtained from the analytical model presented in \cite{Torelli_2018} as a function of U, using the D and $\lambda$ results obtained from PBE+U (PAW) calculations and the $J$ results obtained from DMC (blue triangle) and PBE+U (green circle) using RRKJ pseudopotentials. The magenta rectangle in a) - b) represents the optimal fitted U value of 2.4(1) eV. }
\label{magfig}
\end{center}
\end{figure}

As previously mentioned, we used the geometry obtained with SCAN (FM and U=0) for the subsequent DMC calculations, because the calculated lattice constant (for the 2x2x1 supercell) of 5.696 \AA\space is identical to the experimental value of 5.69(1) \AA. DMC has the zero-variance property which means that as the trial wave function approaches the exact ground state (i.e., the exact nodal surface), the statistical fluctuations in the energy reduce to zero \cite{RevModPhys.73.33}. There have been instances where various sophisticated, often times expensive methods have been used to optimize the nodal surface of the trial wave function \cite{PhysRevB.48.12037,PhysRevB.58.6800,PhysRevE.74.066701,PhysRevLett.104.193001}. However, similar to other DMC studies of correlated magnetic materials \cite{PhysRevX.4.031003,PhysRevMaterials.5.064006,PhysRevMaterials.3.124414,PhysRevMaterials.2.085801,staros}, we used a PBE+U approach where the Hubbard U value was used as a variational parameter to optimize the nodal surface using DMC (for the FM and AFM states of 2D MnO$_2$). The fact that we can determine the optimal U parameter variationally using DMC makes our DMC results more reliable than DFT+U, where in DFT+U the U parameter is usually arbitrarily chosen or fitted to experimental data. For the case of 2D MnO$_2$, where experimental data for properties such as $T_c$ are not yet measured, the DMC determined U value can be used as a fitting parameter for subsequent DFT+U calculations. The results of  these calculations (creating the nodal surface with different U values) are depicted in Fig. \ref{geofig} d), where we observe an energy minimum around U = 2.5 eV for the FM and AFM states. To determine the optimal value of U, we performed a quadratic fit on the FM data depicted in Fig. \ref{geofig} d) and obtained a value of U = 2.4(1) eV. We performed this fit on the FM data rather than the AFM data due to the fact that the error is smaller for the FM data and the deviation from a perfect quadratic fit is minimal (in contrast to the large deviation at U = 4 eV for the AFM data). The range of optimal U (2.4(1) eV) is depicted by the magenta rectangle in Fig. \ref{geofig} d) - e) and Fig. \ref{magfig} a) - b). It is important to note that energies are statistically indistinguishable for the separate FM and AFM configurations from U = 2 - 3.5 eV, demonstrating DMC's weaker dependence on the Hubbard correction. 

The full form of the model spin Hamiltonian \cite{Torelli_2018,Lado_2017} is: 

\begin{equation}
{\cal {H}}= - \left( \sum_{i} D(S_i^z)^2 +\frac{J}{2}\sum_{i,i'}\vec{S}_i\cdot \vec{S}_{i'} + \frac{\lambda}{2}\sum_{i,i'}S_i^z S_{i'}^z \right)
\label{hamiltonian}
\end{equation}
where the sum over $i$ runs over the lattice of Mn atoms and $i'$ runs over the nearest Mn site of atom $i$ due to strong magnetic moment localized on Mn atoms. Long-range interactions are shown to die out in other 2D magnetic materials \cite{Lado_2017}, so we focused solely on the nearest neighbor interactions. The first term in the Hamiltonian describes the easy axis single ion anisotropy (with $z$ chosen as the off-plane direction). The second term is the Heisenberg isotropic exchange and the last term is the anisotropic exchange. The sign convention follows such that $J>0$ favors FM interactions, $D>0$ favors off-plane easy axis and $\lambda=0$ implies completely isotropic exchange.

First we treat Eq. \ref{hamiltonian} classically, describing the spins $\vec{S}$ collinearly as either $S=S^x$ or $S=S^z$. This makes it possible to write the energy of the ground state for 4 possible ground states: i) ferromagnetic off-plane (FM,$z$), ii) antiferromagnetic off-plane (AFM,$z$), iii) ferromagnetic in-plane (FM,$x$), and iv) antiferromagnetic in-plane (AFM,$x$). The corresponding energy equations normalized for the 2x2x1 supercell of 2D MnO$_2$ (4 Mn atoms) are as followed:
\begin{equation}
\textrm{E}_{\textrm{FM},z} = -4S^2D-12S^2(J+\lambda)
\label{fmz}
\end{equation}

\begin{equation}
\textrm{E}_{\textrm{AFM},z} = -4S^2D+4S^2(J+\lambda)
\label{afmz}
\end{equation}

\begin{equation}
\textrm{E}_{\textrm{FM},x} = -12S^2J
\label{fmx}
\end{equation}

\begin{equation}
\textrm{E}_{\textrm{AFM},x} = +4S^2J
\label{afmx}
\end{equation}
where $S=3/2$. We extracted the value for $J$ from our ground state energies (FM and AFM states) calculated with spin-polarized DFT+U and DMC using PBE and RRKJ pseudopotentials. In our DMC and DFT+U calculations (RRKJ), the spin on the Mn atoms is strictly up or down (spin-orbit calculations are currently not available for DMC and RRKJ pseudopotentials do not have an explicit spin-orbit contribution). This excludes us from calculating the energies from Eq. \ref{fmx} and \ref{afmx} using DMC and DFT+U (RRKJ) simulations due to the needed non-collinear magnetism simulations. In addition to these, only the $J+\lambda$ term can be calculated using Eq. \ref{fmz} and \ref{afmz} for DMC and DFT+U (RRKJ). For now, we neglect the anisotropic exchange ($\lambda$) term and calculate the magnetic exchange $J$ parameter from Eq. \ref{fmz} and \ref{afmz}. We will clarify why $\lambda$ can be omitted for 2D MnO$_2$ in upcoming discussions.  

Figure  \ref{geofig} e) depicts the nearest neighbor magnetic exchange parameter ($J$) obtained with DMC and DFT+U as a function of U value (using the PBE functional and RRKJ pseudopotentials). The $J$ parameter calculated with DFT increases linearly as U is increased, signifying the strong dependence of the Hubbard parameter on magnetic exchange at the DFT level. Despite this dependence, for all U values (including U = 0 eV), $J>0$, signifying the FM state as more energetically favorable.  

On the other hand, DMC has a much weaker dependence on the U value used to create the nodal surface (from the separate FM and AFM calculations), which is indicated in Fig. \ref{geofig} e), where we see the $J$ parameter is statistically identical for U = 1 - 3.5 eV. For the optimal value of U = 2.5 eV (as seen in Fig. \ref{geofig} d)), we obtain a value of $J=1.2(5)$ meV. At the DMC level for U = 4 eV (not depicted in Fig. \ref{geofig} e)), we observe a change from FM favorable to AFM favorable ($J=-1.1(7)$ meV),  indicating that we are out of the regime where the Hubbard parameter dependence is weak (see Fig. \ref{geofig} d)). Similar behavior of $J$ with respect to U has been observed in other DMC works such as in Foyevtsova et al. \cite{PhysRevX.4.031003}. 

For explicit calculations of $\lambda$ and $D$, we performed self-consistent spin-orbit PBE+U calculations with PAW pseudopotentials. Specifically in these noncollinear calculations, we rotated the easy axis by 90$^{\circ}$ and calculated the energy difference between the rotated and non-rotated configurations (for FM and AFM states separately). $D$ and $\lambda$ as a function of U are depicted in Fig. \ref{magfig} a). We observe $D$ to be on the order of $\sim$0.1 meV and $\lambda$ to be on the order of $\sim$1 x 10$^{-3}$, which implies that the easy axis single ion anisotropy is much more dominant than the anisotropic symmetric exchange. Since magnetic anisotropy favoring vertical orientation of the spins to the layer does in fact exists for 2D MnO$_2$, we can confirm that FM ordering can exist in 2D for this system. In addition since $D>0$ for all values of U, the spins favor the off-plane easy axis. As seen in Fig. \ref{magfig} a), changing the Hubbard parameter changes the value for $D$ and $\lambda$. $D$ tends to decrease as U is increased and $\lambda$ decreases until U = 3 eV, and then slightly increases from U = 3 - 4 eV. Due to the fact that the value of $\lambda$ is so small comparatively to $D$ and $J$, and the fact that the values of $\lambda$ for 2D MnO$_2$ are roughly two orders of magnitude smaller than 2D CrI$_3$ \cite{Lado_2017}, we can infer that the contribution of $\lambda$ is negligible and the exchange for 2D MnO$_2$ is completely isotropic. In addition, this is the reason why we can safely avoid $\lambda$ for determining the $J$ parameter from DMC and DFT+U simulations (RRKJ potentials) (Eq. \ref{fmz} and \ref{afmz}). We can also infer that the slight increase in $\lambda$ after U = 3 eV is simply an artifact of the calculation, since it is numerically difficult to determine $\lambda$ due to its small magnitude. At the value of U = 2.5 eV, the Hubbard value we obtained the optimal DMC nodal surface for, we calculated $D=0.118$ meV and $\lambda=\sim 2$ x 10$^{-3}$ meV. 


From our calculations of $J$, $D$ and $\lambda$, it is possible to estimate the critical temperature using the method outlined in Torelli and Olsen \cite{Torelli_2018}. In this method, the results from classical Monte Carlo and Random Phase Approximation (RPA) simulations were used to derive a simple analytical expression for critical temperature ($T_c$) that depends on lattice symmetry and is a function of the exchange coupling constants ($J$, $D$, $\lambda$).  Such an expression was derived to significantly simplify the theoretical search for new 2D magnetic materials with high critical temperatures. By fitting the classical simulations, an analytical function for $T_c$ takes the following form.
\begin{equation}
T_c=T_c^{\textrm{Ising}}f(x)
\label{tc}
\end{equation}
with 
\begin{equation}
f(x)=\textrm{tanh}^{1/4}\left[\frac{6}{N_{nn}}\text{log}(1+\gamma x)\right]
\label{fx}
\end{equation}
where $\gamma=0.033$ and $N_{nn}$ is the number of nearest neighbors. $T_c^{\textrm{Ising}}$ is the critical temperature for the corresponding Ising model, which takes the form $T_c^{\textrm{Ising}}=S^2 J \tilde{T}_c/k_B$, where $\tilde{T}_c$ is the fitted dimensionless critical temperature (3.64 for trigonal lattice). When single ion anisotropy and anisotropic exchange are both present, $x=\Delta/J(2S-1)$, where $\Delta$ is the spin gap  that takes the form:
\begin{equation}
\Delta=D(2S-1)+\lambda S N_{nn}
\label{delta}
\end{equation}

Figure \ref{magfig} b) depicts the critical temperature as a function of U value. The green circles represent $T_c$ calculated from $J$ determined with strictly PBE+U (RRKJ potentials) and $D$ and $\lambda$ determined from spin-orbit PBE+U calculations (PAW potentials). The blue triangles and corresponding error bars represent $T_c$ calculated from $J$ determined with DMC and also $D$ and $\lambda$ determined from spin-orbit PBE+U calculations (PAW potentials). As seen in Fig. \ref{magfig} b), $T_c$ calculated with strictly DFT (green circles) linearly increases as a function of U, similarly to how $J$ determined with DFT increases as a function of U (see Fig.\ref{geofig} e)). Although the inclusion of MA decreases $T_c$ by $f(x)=1/5$ (see Eq. \ref{tc} - \ref{delta}), the variation of $T_c$ (strictly DFT) with respect to U is mainly dominated by the larger changes in $J$ rather than the smaller changes in $D$ and $\lambda$ (see Fig. \ref{magfig} a) and Fig. \ref{geofig} e)). 

In order to make a more systematic comparison between higher order methods and provide an additional benchmark, we also calculated the FM, AFM and MA energies with the screened HSE06 functional (using PAW potentials), which includes a portion of exact Fock exchange and can estimate magnetic properties with high accuracy \cite{doi:10.1063/1.1564060}. With this method, we obtain a $J$ value of 1.79 meV, a $D$ value of 0.162 meV and a $\lambda$ value of $3$ x 10$^{-3}$ meV. Using these values in conjunction with the Torelli and Olsen model, we calculated a $T_c$ of 32.9 K. This can be directly compared to the the PBE+U (at the optimal U value of 2.5 eV) calculated $T_c$ value of 29.2 K, where it is in close agreement with the HSE06 obtained value (tabulated results are given in Table S1 for comparison). Determining $T_c$ with the $J$ extracted from DMC simulations allows us to place an upper and lower bound on the result (see Fig. \ref{magfig} b)). Using the optimal nodal surface obtained with U = 2.5 eV for our DMC calculated $J$ (and subsequently the $D$ and $\lambda$ values obtained for U = 2.5 eV), we calculated a $T_c$ value of 22.2 $\pm$ 6.6 K. If the HSE06 calculated values for magnetic anisotropy parameters are used instead of using the $D$ and $\lambda$ from PBE+U (U = 2.5 eV), we obtain a $T_c$ value of 24.1 $\pm$ 6.6 K. Although this critical temperature is far below room temperature (and approximately half of the $T_c$ of 45 K measured for 2D CrI$_3$), it has been demonstrated that $T_c$ can be increased by applying strain \cite{fm-mno2} or by placing the monolayer on a substrate \cite{vse2}. To illustrate that $J$ is the driving force behind the change in $T_c$ (in comparison to $D$ and $\lambda$), we calculated $T_c$ with a fixed value of $J$ (obtained from calculations at U = 2.5 eV) and varied $D$ and $\lambda$ as a function of U, depicted in Fig. S3. From Fig. S3 we observe that $T_c$ remains constant for DFT and DMC for a fixed $J$ value while $D$ and $\lambda$ vary as a function of U.

\begin{figure}
\begin{center}
\includegraphics[width=8.5cm]{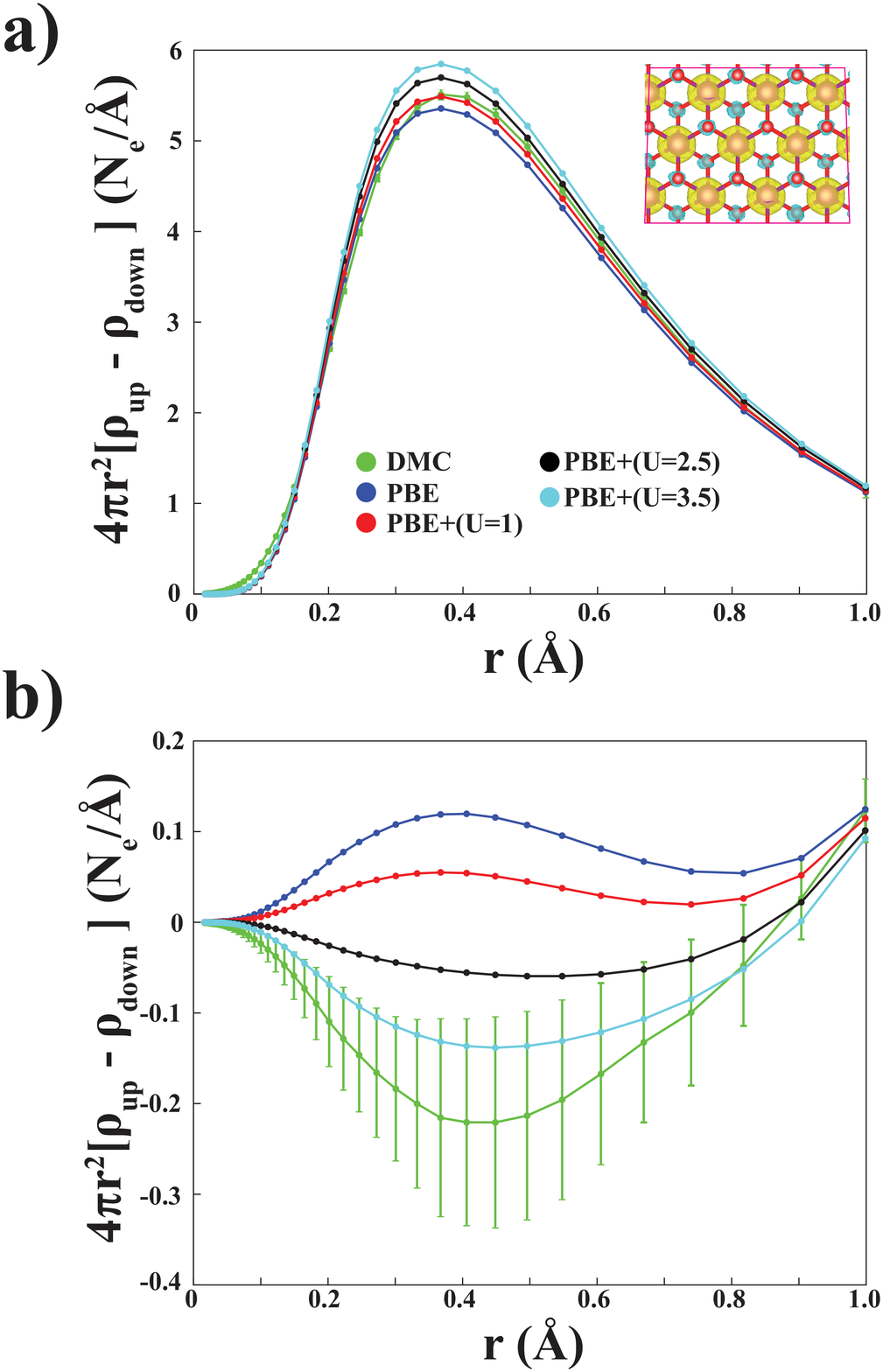}
\caption{The spin density ($\rho_{up} - \rho_{down}$) calculated with DMC and PBE+U (U = 0, 1, 2.5, 3.5 eV) for a) Mn and b) O. The inset of a) depicts the spin isosurface density of 2D MnO$_2$ where the isosurface value was set to 5 x 10$^{-5}$ e/\AA$^{3}$.}
\label{spindensity}
\end{center}
\end{figure}

As an additional theoretical benchmark, we extracted the total charge density from our DMC calculations from Eq. \ref{rho1} using optimal U value from Fig. \ref{geofig} d). From this extracted total charge density, we determined the spin density, ($\rho_{up} - \rho_{down}$). The inset of Fig. \ref{spindensity} a) depicts the spin isosurface density calculated with DMC of 2D MnO$_2$. Using this many-electron approach, we observe that the Mn atoms are highly spin-polarized while the O atoms are slightly polarized anti-parallel with respect to Mn atoms. To explore these properties further and benchmark with various DFT methods, we determined the spatial variations in total charge density (Fig. S4 a) - b)) and spin density (Fig. \ref{spindensity} a) - b)) by plotting the radial averaged densities as a function of distance for Mn and O separately. For Fig. S4 and \ref{spindensity}, DMC is benchmarked with DFT+U (U = 0, 1, 2.5, 3.5) and RRKJ pseudopotentials are used for all calculations. Figure S4 a) depicts the total radial density for Mn. We observe that while all DFT+U results are almost indistinguishable, they significantly overestimate the density of Mn (especially around the peak at r = 0.4 \AA). For O, the differences for total radial density between DMC and DFT+U is less apparent, with DMC slightly underestimating the density from r = 0 to r = 0.4 \AA\space and slightly overestimating after r = 0.4 \AA. This larger discrepancy in the Mn atom near the radial density peak (peak of $d$ orbital) is due to the fact that DFT functionals tend to unsuccessfully capture 3$d$ orbitals. 

\begin{table}[]
\caption{\label{tab:moment}
The site-averaged atomic magnetic moments of Mn and O estimated by integrating the spin density for DMC and PBE+U results. }
\begin{tabular}{llr}

\hline
\hline
Method & M$_{\textrm{Mn}}$ ($\mu_B$)  & M$_{\textrm{O}}$ ($\mu_B$)\\

\hline
DMC    & 2.77(1)        & -0.10(2)                       \\
PBE, U=0   & 2.69        & 0.06                        \\
PBE, U=1       & 2.76       & 0.03                      \\
PBE, U=2.5       & 2.86       & -0.03                        \\
PBE, U=3.5       & 2.93        & -0.07                        \\

\hline
\hline
\end{tabular}
\end{table}

Although there are sizable differences for the total charge densities between PBE+U and DMC, it has been reported that various DFT methods generally give a more accurate description of spin density than the total density \cite{PhysRevMaterials.1.065408}. Figure \ref{spindensity} a) depicts the radial spin density of Mn. From this figure, we observe a much closer agreement between PBE+U and DMC than for the total charge density for Mn, where the exact same shape of the curve is observed for all values of U. This demonstrates that a high quality spin density for Mn in MnO$_2$ can be calculated despite the different U values. In contrast to Mn, the radial spin density of O is depicted in Fig. \ref{spindensity} b). From our DMC result, we confirm that the O atoms are slightly polarized anti-parallel with respect to Mn atoms. From our PBE+U benchmarking, we observe that this anti-parallel polarization is completely dependent on which value of U is used. For U = 0, 1 eV the radial curve is concave up while after U = 2.5 eV, the curve is concave down, which indicates that after an effective value of U, the polarization of O atoms is corrected. These results are in accordance with our energetic results, where we determined the value of U = 2.5 eV to yield the lowest DMC energy, with U = 3.5 eV being statistically identical (see Fig. \ref{geofig}). We went one step further and estimated the site-averaged atomic magnetic moments per Mn and O by integrating the spin densities depicted in Fig. \ref{spindensity}. The tabulated magnetic moments are presented in Table \ref{tab:moment}, where the results match the trends depicted in Fig. \ref{spindensity}. In contrast to the values presented in Table \ref{tab:moment}, HSE06 (PAW) predicts a magnetic moment of 2.97 $\mu_B$ for Mn and -0.02 $\mu_B$ for O, indicating that the HSE06 Mn magnetic moment is closest to PBE+U = 3.5 and the HSE06 O magnetic moment is closest to PBE+U = 2.5. By integrating the spin densities, we have a clear picture of how the magnetization of each ion changes with respect to the electronic structure method used. Our total charge density and spin density results and magnetic moment estimates serve as an ultimate many-body theoretical benchmark for the magnetic properties of 2D MnO$_2$ and give insight on how to assess the accuracy of DFT calculations when various values of U are employed.

\section{\label{sec:conclusion}Conclusion}

To resolve the discrepancies that arise in DFT calculations that have a strong dependence on density functional and Hubbard parameter (U), we employed DMC to calculate the energetic and magnetic properties of monolayer MnO$_2$. From these calculations, we found that the FM phase is more energetically favorable than the AFM phase and found the optimal U value that yields the lowest total energy to be 2.4(1) eV. By taking the difference of the FM and AFM energies calculated with DMC, we were able to estimate a Heisenberg isotropic exchange parameter ($J$) of 1.2(5) meV at the optimal U value. Using spin-orbit DFT+U (PAW method), we calculated the single ion anisotropy ($D$) and the anisotropic symmetric exchange ($\lambda$). By combining the DMC results for $J$ and DFT+U results for $D$ and $\lambda$ obtained at U = 2.4(1), we estimated the upper bound on $T_c$ to be 28.8 K. Additionally, we extracted the spin-density isosurfaces and the radial averaged spin density for Mn and O atoms separately from our DMC results and provide a detailed comparison with DFT+U. Our findings demonstrate the success of DMC being applied to a 2D magnetic system and provide an ultimate theoretical benchmark that will aid in guiding experimentalists in synthesizing and characterizing 2D magnets.

\begin{acknowledgments}
This work was supported by the National Science Foundation through the Division of Materials Research under NSF DMR-1726213. The authors would like to thank Dr. Yelda Kadioglu for fruitful discussions.
\end{acknowledgments}

\section*{References}
\bibliography{acs-acschemso}

\end{document}


\maketitle


\begin{figure}
\begin{center}
\includegraphics[width=8.5cm]{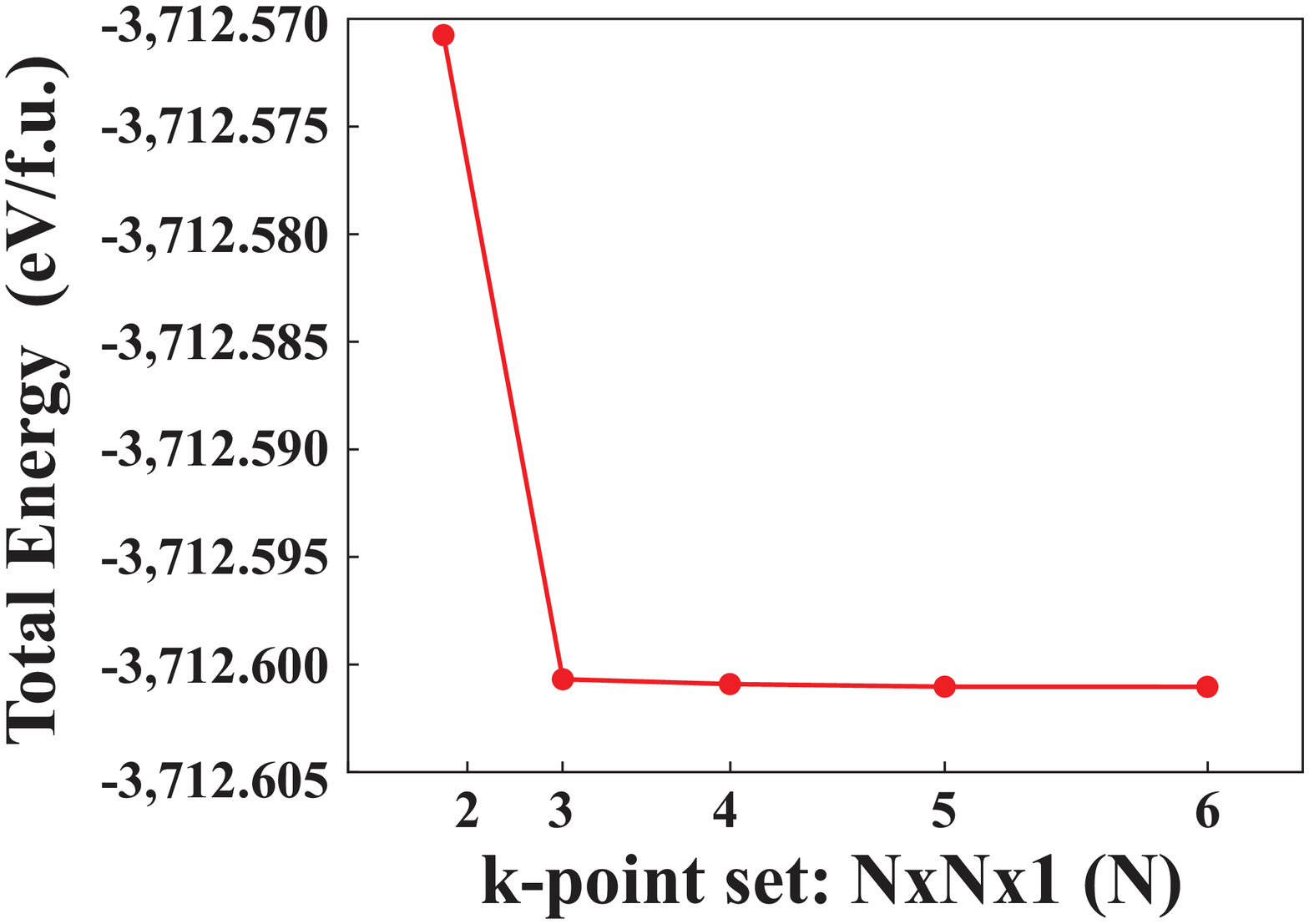}
\caption{The total energy per formula unit of the 2x2x1 supercell of 2D MnO$_2$ (12 atoms) as a function of k-point grid
for the norm-conserving RRKJ pseudopotentials calculated with DFT (PBE).
The results show a converged k-point grid of 3x3x1. The number of k-points was scaled appropriately to obtain the converged grid depending on the supercell size and shape for all DFT and QMC calculations.}
\label{appendix-1}
\end{center}
\end{figure}

\begin{figure}
\begin{center}
\includegraphics[width=8.5cm]{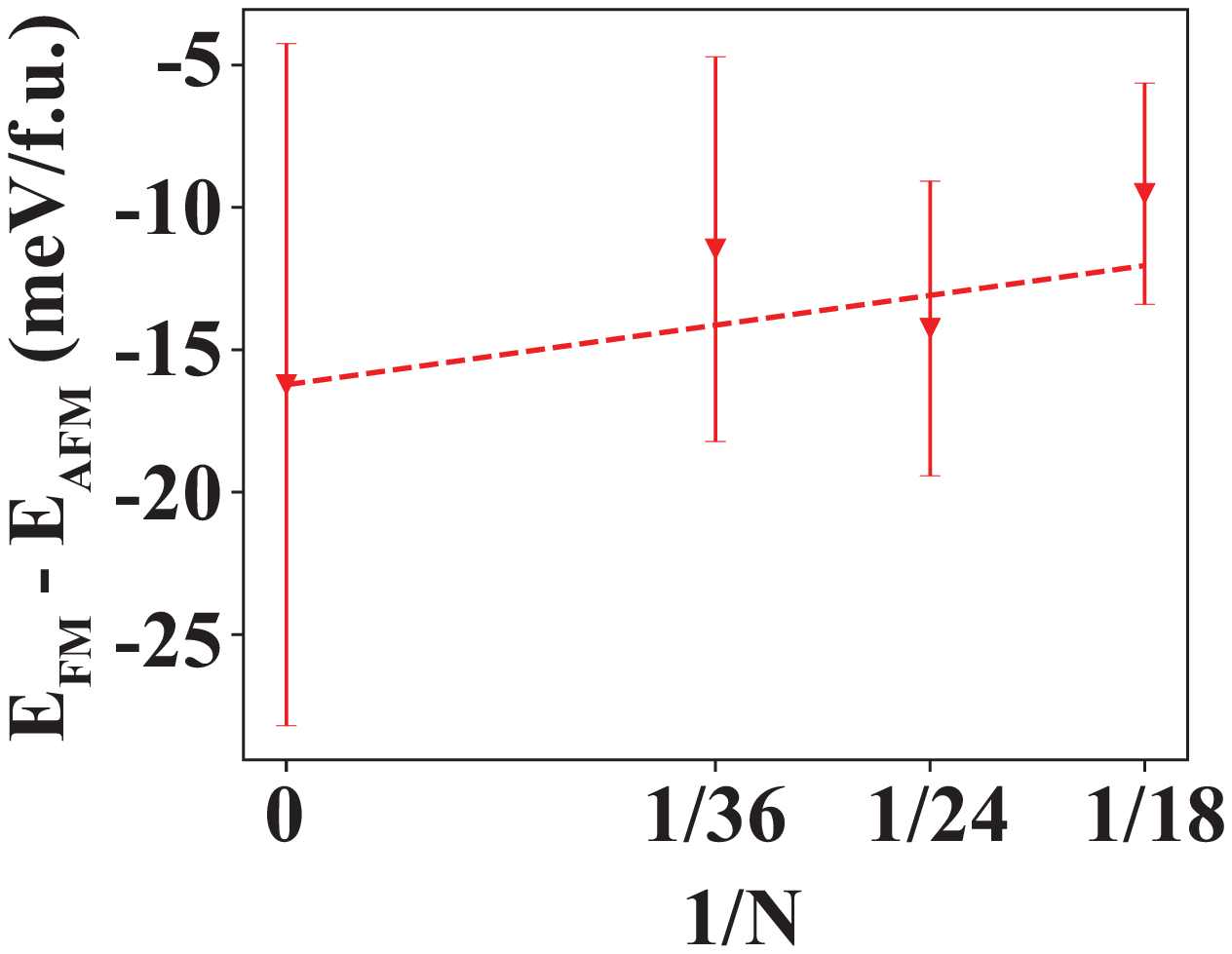}
\caption{The energy difference of the FM and AFM configurations of MnO$_2$ (per f.u.) as a function of supercell sizes (N = 18, 24, and 36 atoms), demonstrating finite size convergence in supercells smaller than those used in this study (36 atom).   }
\label{fs_correct}
\end{center}
\end{figure}

\begin{figure}
\begin{center}
\includegraphics[width=7.5cm]{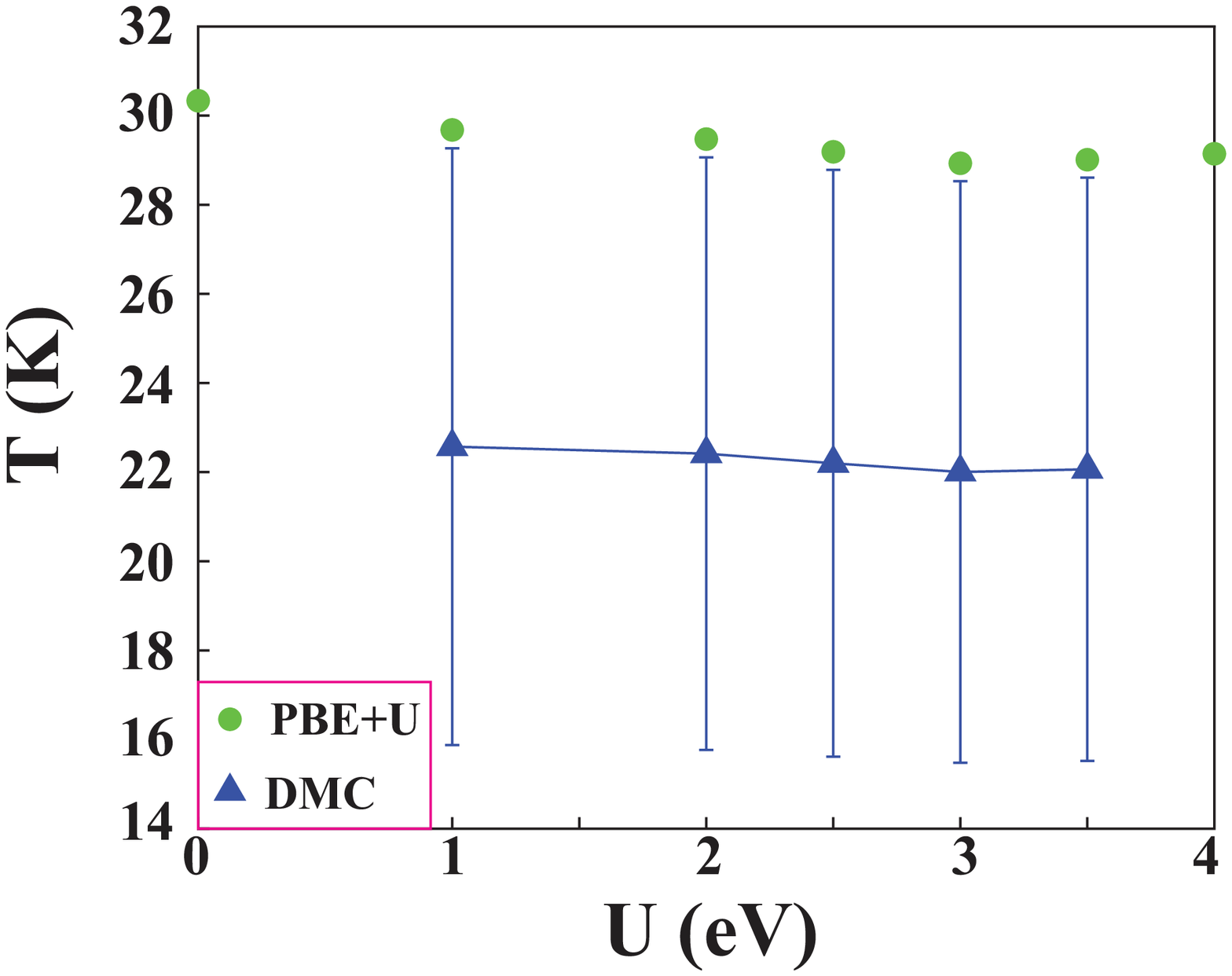}
\caption{ The estimated critical temperature obtained using DMC (blue triangle) and PBE+U (green circle). In this case, $J$ is a fixed value obtained from the DMC and PBE+U results for U = 2.5 eV while D and $\lambda$ vary as a function of U. }
\label{appendix-3}
\end{center}
\end{figure}

\begin{figure}
\begin{center}
\includegraphics[width=7.5cm]{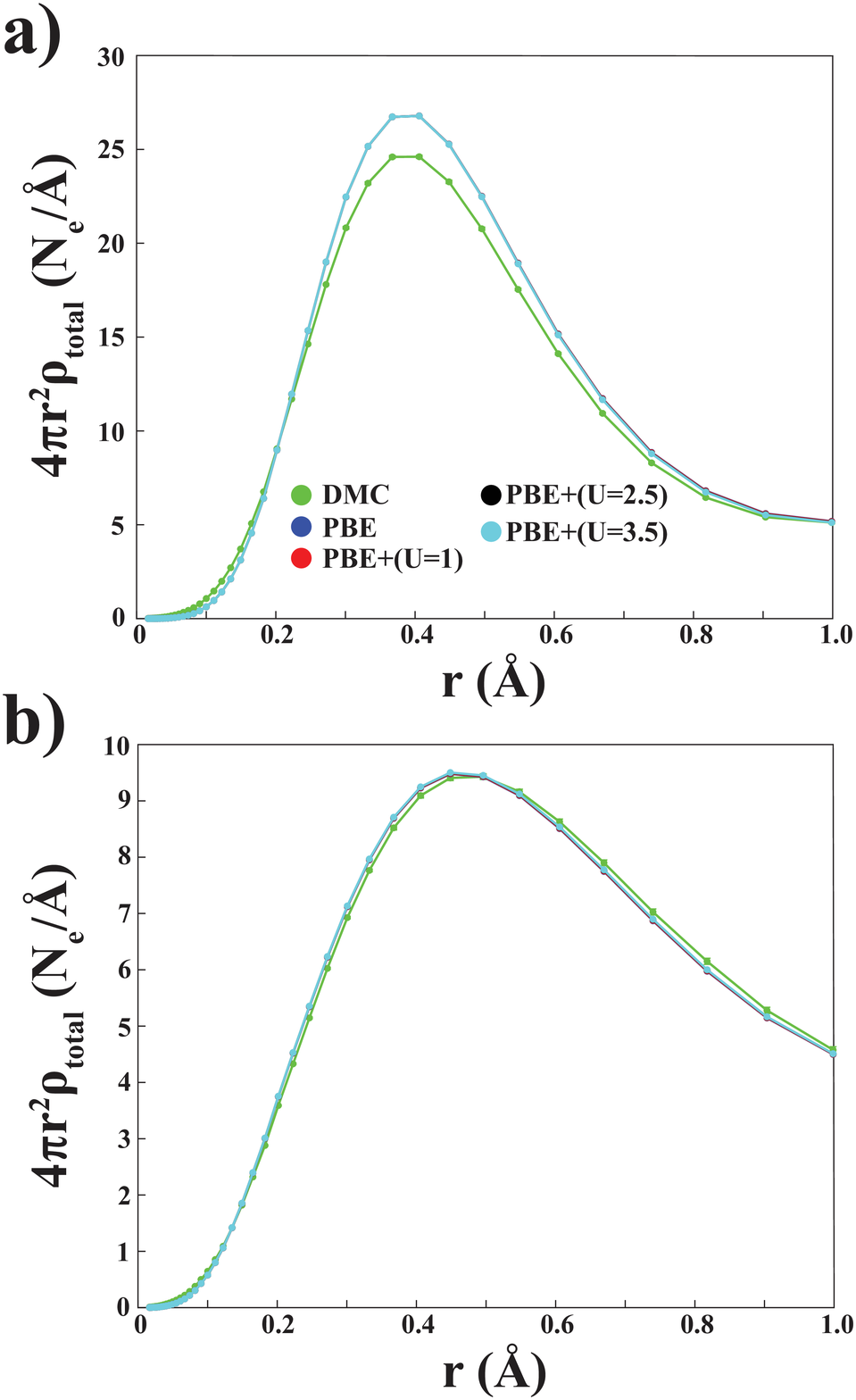}
\caption{The total radial charge density calculated with DMC and PBE+U (U = 0, 1, 2.5, 3.5 eV) for a) Mn and b) O. All PBE+U data points are indistinguishable from each other. }
\label{appendix-4}
\end{center}
\end{figure}

\begin{table}[]
\caption{\label{tab:moment}
Tabulated results for $J$, $D$, $\lambda$, and $T_c$ calculated with DMC (RRKJ at optimal nodal surface of U = 2.5 eV), PBE+U (RRKJ for $J$, PAW for $D$ and $\lambda$ at optimal nodal surface of U = 2.5 eV), and HSE06 (PAW). $^\dagger$ represents the $T_c$ value obtained from $J$ calculated with DMC and $D$ and $\lambda$ calculated from PBE+U while $^\star$ represents the $T_c$ value obtained from $J$ calculated with DMC and $D$ and $\lambda$ calculated from HSE06.  }
\begin{tabular}{llllr}

\hline
\hline
Method & $J$ & $D$ & $\lambda$ & $T_c$ \\
& (meV) & (meV) & (meV) & (K) \\
\hline
DMC, U=2.5     & 1.2(5)   &     -       & - &   22.2(6.6)$^\dagger$               \\

& & & & 24.1(6.6)$^\star$\\
PBE, U=2.5       & 1.71       &  0.116   &  0.002 &     29.19                \\
HSE06 & 1.79 & 0.162 & 0.003 & 32.93 \\

\hline
\hline
\end{tabular}
\end{table}